\documentclass[preprint,showpacs,preprintnumbers,superscriptaddress,amsmath,amssymb]{revtex4}
\usepackage{graphicx}
\usepackage{dcolumn}
\usepackage{bm}
\usepackage{hyperref}

\begin{document}
\title{Temperature dependent local structure of NdFeAsO$_{1-x}$F$_x$ system using arsenic $K$-edge 
extended x-ray absorption fine structure}
\author{Boby Joseph}
\affiliation{Dipartimento di Chimica - Sezione di Chimica Fisica, INSTM (UdR Pavia), Universit\`{a} di Pavia, Viale Taramelli 16, 27100 Pavia, Italy}
\author{Antonella Iadecola}
\affiliation{Dipartimento di Fisica, Universit\`{a} di Roma ``La 
Sapienza", P. le Aldo Moro 2, 00185 Roma, Italy}
\author{Lorenzo Malavasi}
\affiliation{Dipartimento di Chimica - Sezione di Chimica Fisica, INSTM (UdR Pavia), Universit\`{a} di Pavia, Viale Taramelli 16, 27100 Pavia, Italy}
\author{Naurang L. Saini}
\affiliation{Dipartimento di Fisica, Universit\`{a} di Roma ``La 
Sapienza", P. le Aldo Moro 2, 00185 Roma, Italy}


\begin{abstract}
Local structure of NdFeAsO$_{1-x}$F$_{x}$ ($x$=0.0, 0.05, 0.15 and 0.18) high temperature 
iron pnictide superconductor system is studied using arsenic $K$-edge extended x-ray
absorption fine structure measurements as a function of temperature.
Fe-As bondlength shows only a weak temperature and F-substitution
dependence, consistent with the strong covalent nature of this bond.
The temperature dependence of the mean-square relative-displacements
of the Fe-As bondlength are well described by the correlated-Einstein
model for all the samples, but with different Einstein-temperatures
for the superconducting and non-superconducting samples.  
The results indicate distinct local Fe-As lattice dynamics in the
superconducting and non-superconducting iron-pnictide systems.\\
Journal reference : Journal of Physics: Condensed Matter {\href{http://iopscience.iop.org/0953-8984/23/26/265701/}{23 (2011) 256701}}

\end{abstract}

\pacs{74.70.Xa;74.81.-g; 61.05.cj; 74.62.Bf}

\maketitle

\section{INTRODUCTION}
The role of structural topology in the magnetic and superconducting
properties of the newly discovered iron-based superconductors (FeSC)
is one of the active themes of investigation
\cite{rev_greene,rev_lumsden,rev_hosono} focussing on finding the
mechanisms of the unconventional superconductivity.  Several studies
have shown unprecedented sensitivity of Fermi surface topology to the
anion (pnictogen or chalcogen) height above the Fe-plane in these
materials.  The anion height is seen to influence the density of
states near the Fermi energy and influence the electron-pairing
properties of the FeSC
\cite{VildoPRB,KurokiPRL,CHLee,anion_ht_takano,egami,Miyake_ab_inito,Mazin_Johannes}.
In addition, local structural studies have clearly pointed out the importance of 
the local order-disorder in
the spacer layer in the FeSC \cite{XANES_REOFeAs,XANES_As_REOFeAs}.
The structurally simplest systems among the FeSC, the
FeTe$_{1-x}$Se$_x$, is found to have lower local structural symmetry 
than the average crystallographic symmetry, with the Se and Te occupying
distinct sites \cite{11-EXAFS,11-PDF}. This local inhomogenity is found to
have a direct consequence on the electronic properties of the system \cite{11-XAS}. 
All these point to the importance of the diverging local structure 
from the average one, putting the FeSC on the same platform of cuprates, 
revealing lattice anomalies and local inhomogeneities \cite{loc_str_G4}, 
closely related with the superconducting properties.  
Some recent EXAFS studies on F-doped LaFeAsO and SmFeAsO have shown a 
weak anomaly in the Fe-As bondlength fluctuations close to the 
superconducting transition \cite{La1111-EXAFS,Sm1111-EXAFS}.  
However, such anomalies are found  to be much weaker, 
compared to those observed in the cuprates \cite{loc_str_G4}.

Here, we have used extended x-ray absorption fine structure (EXAFS), a
fast and site-specific experimental tool \cite{Konings,book2}, to probe
systematically the local structure of NdFeAsO$_{1-x}$F$_{x}$ oxypnictides 
as a function of temperature and charge density.  Arsenic $K$-edge EXAFS has been used
to retrieve direct information on the Fe-As bondlengths in
superconducting (x=0.15 and 0.18) and non-superconducting (x=0.0 and
0.05) samples.  The bondlengths show weak temperature and F-doping
dependence, consistent with the strong covalent nature of the Fe-As
bonds.  The corresponding mean square relative displacements (MSRDs),
well described by the correlated Einstein model, reveal different
Einstein-temperatures ($\Theta_{E}$) for the superconducting
($\Theta_{E}$=348$\pm$12 K) and non-superconducting
($\Theta_{E}$=326$\pm$12 K) samples.  The superconducting sample with
maximum T$_{c}$ ($x$ = 0.18) appears to show a weak temperature
dependent anomaly in the Fe-As MSRDs. 
However, the anomaly is well within the experimental uncertainties,
similar to case of SmFeAsO$_{0.931}$F$_{0.069}$.  These results
indicate that there exists, distinct lattice dynamics in the
superconducting and non-superconducting systems, albeit the
differences are smaller.

\begin{figure}
\includegraphics[width=10.0 cm]{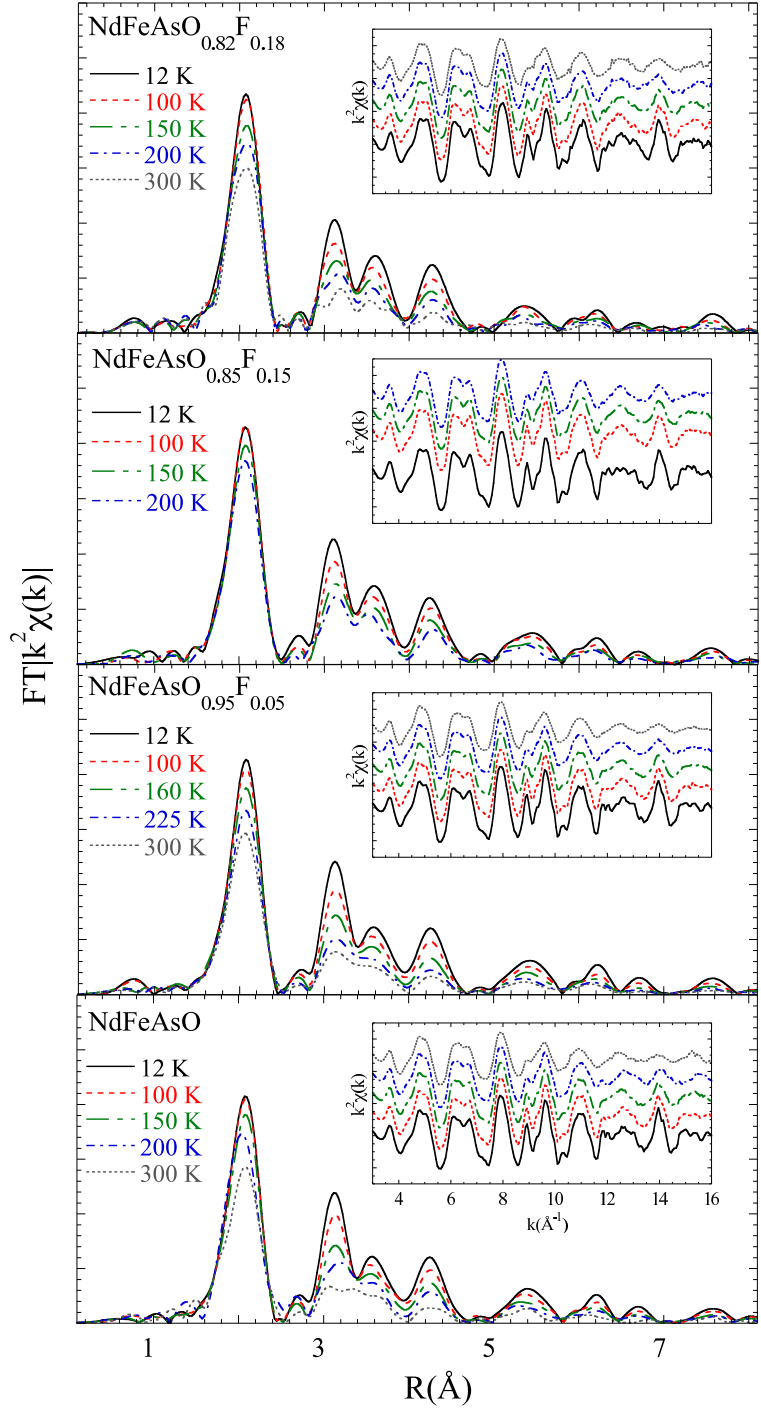}
\caption{\label{fig:epsart} 
Fourier transform (FT) magnitudes of the arsenic $K$-edge EXAFS oscillations for the 
NdFeAsO$_{1-x}$F$_x$ (x=0.0, 0.05, 0.15 and 0.18). Corresponding EXAFS oscillations are 
shown in the insets.  The FTs are not corrected for the phase shifts, thus represent raw 
experimental data.  A systematic temperature dependence is evident from the gradual 
change in the FT intensities.}
\end{figure}
\section{EXPERIMENTAL METHODS}

Temperature dependent x-ray absorption measurements were performed in
transmission mode on powder samples of NdFeAsO$_{1-X}$F$_x$ ($x$=0.0,
0.05, 0.15 and 0.18; the $x$-values given corresponds to the nominal
composition, the actual values are lower than the nominal values for
higher $x$ \cite{malavasi_jacs}), at the BM26A beamline
\cite{BM26A_ESRF} of the European synchrotron radiation facility,
Grenoble (France).  The synchrotron light emitted by a bending
magnetic source was monochromatized by a double crystal Si(111)
monochromator.  The samples, synthesized by conventional solid state
reaction method \cite{malavasi_jacs}, were characterized for their
structural, magnetic and superconducting properties prior to the
experimental run.  While the samples $x$=0.0 and 0.05 are
non-superconducting, showing tetragonal to orthorhombic structural
phase transition, the samples with $x$=0.15 and 0.18 are
superconducting with T$_{c}$=$\sim$20 K and $\sim$40 K respectively,
without any evidence of the structural phase transition
\cite{malavasi_jacs}.  
For the temperature dependent measurements (15-300 K), a continuous flow He cryostat 
with a temperature control within an accuracy of $\pm$1 K was used.
A minimum of two scans (with high signal to noise ratio) were measured
at each temperature to make sure the reproducibility.  Standard
procedure was used to extract the EXAFS oscillations from the
absorption spectrum \cite{Konings,book2}.

\section{RESULTS AND DISCUSSIONS}

\begin{figure}
\includegraphics[width=10 cm]{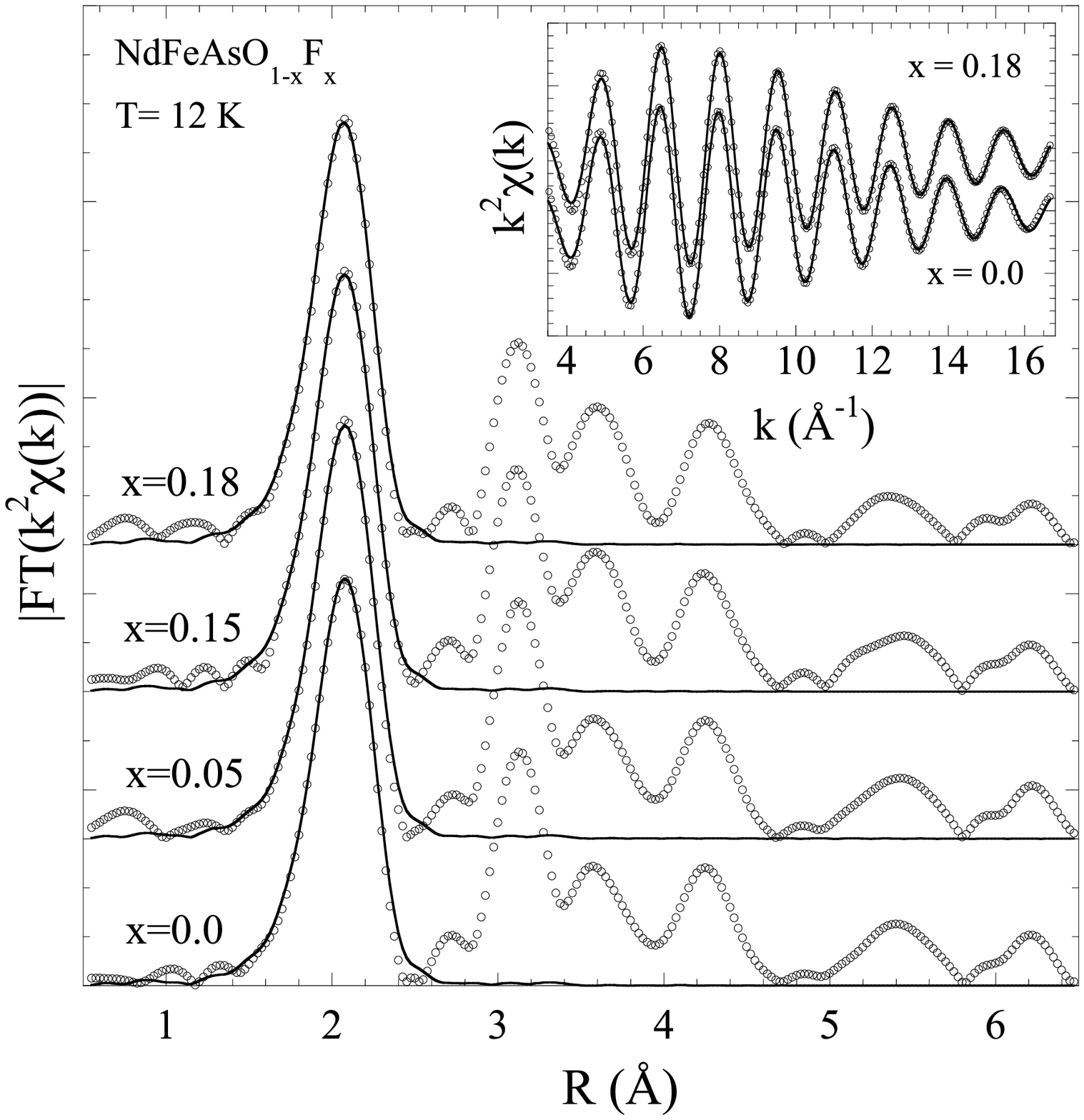}
\caption{\label{fig:epsart}
Fourier transforms of the arsenic $K$-edge EXAFS oscillations measured (symbols) on 
NdFeAsO$_{1-x}$F$_x$ (x=0.0, 0.05, 0.15 and 0.18) at low temperature (12 K), together 
with a single shell fit (solid line).  
The filtered EXAFS oscillations corresponding to the first shell (As-Fe pairs) together with
the model fits (solid lines) for $x$=0.0 and $x$=0.18 samples at 12 K are shown in the inset.}
\end{figure}
Figure 1 shows Fourier transform (FT) magnitudes of the arsenic
$K$-edge EXAFS oscillations (the insets show the EXAFS oscillations),
measured on NdFeAsO$_{1-x}$F$_x$ (for $x$=0.0, 0.05, 0.15 and 0.18)
samples, providing partial atomic distribution around the arsenic
atoms.  At room temperature, all the samples studied have tetragonal
structure (space group P4/nmm) \cite{malavasi_jacs}.  The $x$=0.0 and
0.05 samples show structural phase transition from tetragonal to
orthorhombic phase below $\sim$ 130 K \cite{malavasi_jacs}.  There are
four Fe atoms as the near neighbors of arsenic at a distance $\sim$
2.39 \AA$ $ (main peak at $\sim$2 \AA$ $).  The next nearest
neighbors are four Nd atoms at a distance $\sim$ 3.28 \AA$ $ and four
oxygen atoms at a distance $\sim$ 3.51 \AA$ $, followed by eight As
atoms at $\sim$ 3.9 \AA. Contributions of these distant shells appear
mixed, giving a multiple structured peak at $\sim$2.5-4.5\AA$ $ (Fig.
1).  Apparently, the atomic distribution around the arsenic is similar
in all the samples, as evident from the FT as well the EXAFS
oscillations.

In the NdFeAsO$_{1-x}$F$_x$ system, the arsenic atoms have the nearest
neighbors as Fe atoms, and their contributions to the arsenic
$K$-edge EXAFS is well separated from all other shell contributions.
This makes the arsenic $K$-edge EXAFS data highly suitable for
extracting quantitative information on the Fe-As bondlengths and the
related mean square relative displacements (MSRDs).  Here we have
exploited this possibility and performed single shell model fits to
the EXAFS oscillations due to the Fe-As bondlengths.

The EXAFS amplitude depends on several factors and is given by the
following general equation \cite{Konings,book2}:
\begin{equation}
\chi(k)= \sum_{i}\frac{N_{i}S_{0}^{2}}{kR_{i}^{2}}f_{i}(k,R_{i})
e^{-\frac{2R_{i}}{\lambda}} e^{-2k^{2}\sigma_{i}^{2}}
sin[2kR_{i}+\delta_{i}(k)]\nonumber
\end{equation}
where N$_{i}$ is the number of neighboring atoms at a distance
R$_{i}$, S$_{0}^{2}$ is the passive electrons reduction factor,
f$_{i}$(k,R$_{i}$) is the backscattering amplitude, $\lambda$ is the
photoelectron mean free path, $\delta_{i}$ is the phase shift, and
$\sigma_{i}^{2}$ is the correlated Debye-Waller (DW) factor, measuring
the mean square relative displacements (MSRDs) of the
photoabsorber-backscatterer pairs.  Apart from these variables, the
photoelectron energy origin, E$_{0}$, is another input needed for the
modeling of the EXAFS. In the present case, we have used a
single-shell EXAFS-modeling \cite{Konings} to extract the Fe-As bond
correlations.  For such a single-shell analysis, we have used the
WINXAS package \cite{winxas}, with backscattering amplitudes and phase
shifts calculated using FEFF \cite{Feff} with the crystal structure
data from diffraction \cite{malavasi_jacs} as input.  In the
single-shell least-square fit, the number of independent data points,
N$_{ind}\sim$(2$\Delta$k$\Delta$R)/$\pi$ \cite{Konings} were about 13
($\Delta$k=14 \AA$^{-1}$ (k=3-17\AA$^{-1}$) and $\Delta$R=1.5 \AA),
but, we used only two parameters, the radial distance R$_{i}$ and the
corresponding MSRD $\sigma_{i}^{2}$, as the refinable variables,
fixing all other parameters (S$_{0}^{2}$=1), to obtain reliable
information on the Fe-As bond correlations. 
The uncertainities in the derived two paramters, R and $\sigma^2$, were 
estimated by standard EXAFS method \cite{book2}. 
The canonical approch to error estimation is to determine the region around
the best fit that contains the true value with a certain probability $\beta$.
The projection of that volume onto an axis corresponding to a parameter
gives the paramter errors, for a chosen value of $\beta$.
In general this uncertainities dependent on the experimental as well as EXAFS 
data extraction procedures, in addition to the uncertainties coming from the 
statistical $\chi^2$-procedure. 
In the present case, due to the adoption of identical experimental conditions 
and EXAFS data extraction procedures, the first kind of uncertainites are minimum.
However, we have set the error bars on derived parameters, R and $\sigma^2$, 
to two times the highest uncertainity estimated. This is to underline the fact 
that we are interested in discussing the relative changes rather than the absolute 
values of these paramters.  
Figure 2 shows the single shell model fit in the real and k-space at 
12 K for samples with different $x$.  Here it should be mentioned that 
the samples used in the present study are phase pure with the amount 
of impurities below the sensitivity of the x-ray diffraction.  
The main impurity phase present in the doped samples is NdOF phase, 
which showed a F-doping
dependence \cite{malavasi_jacs}.  However, in the present case, this
impurity phase no influence on the arsenic $K$-edge EXAFS data.  The
impurities which possibly can interfere with the present EXAFS
results, like, the FeAs phases, NdAs phases were less than 1\%
\cite{malavasi_jacs}, thus ruling out any contribution from such
impurity phases in the data presented here.

\begin{figure}
\includegraphics[width=10.0 cm]{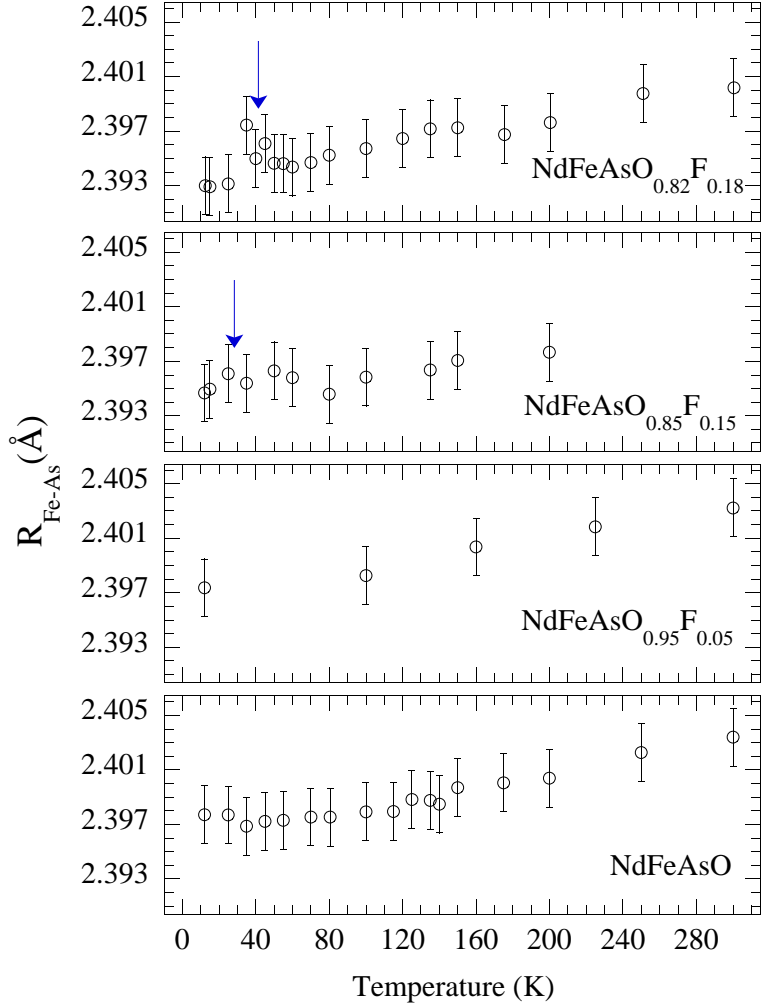}
\caption{\label{fig:epsart} 
Temperature dependence of the Fe-As bond distance for the
NdFeAsO$_{1-x}$F$_x$ system.  The Fe-As bond-length show only weak
doping and temperature dependence, indicating covalent nature of the
bond.  However, there appears a small anomaly (arrow marks in the
upper panels) in the superconducting samples close to the transition
temperature, with no such changes evidenced in the non-superconducting
compounds.}
\end{figure}

Figure 3 shows temperature dependence of the Fe-As bond distances
determined from the single shell analysis of the arsenic $K$-edge
EXAFS data.  The Fe-As distance in all the samples are very similar,
around 2.39 \AA. The over-all temperature dependence of the distance
also looks very similar.  However, one can notice a small change in
the superconducting samples somewhat close to the transition
temperature, with no such changes evidenced in the parent compound.
An earlier temperature dependent x-ray diffraction study on the
NdFeAsO$_{0.85}$ superconductor has showed an abrupt change in the
Fe-As distance around T$_c$ \cite{anomaly_FeAs}.  Present results also
show that the Fe-As distance tend to change near the T$_c$ for the
superconducting (x=0.15, 0.18) samples (Fig.  3 upper panels).  But,
we do not see significant changes in the Fe-As bondlengths or
corresponding MSRD across the structural phase transition for the
parent compound, which is consistent with earlier EXAFS studies on
LaFeAsO \cite{La1111-EXAFS,La1111-EXAFS-2}, SmFeAsO
\cite{Sm1111-EXAFS} and BaFe$_2$As$_2$ \cite{Ba122-EXAFS}, showing no
change in the Fe-As bondlength and corresponding MSRD across the
structural phase transition.  As shown in a recent diffraction study
of the REFeAsO (RE stands for rare-earth) system, data from single
crystals are important in for a better understanding of structural
phase transition properties \cite{xrd-REFeAsO}.  For example, a recent
study on the NdFeAsO single crystal revealed two low-temperature phase
transitions in addition to the tetragonal-to-orthorhombic transition
at T$_S$ $\sim$142 K \cite{3SPT_Nd} indicating the importance of
single crystals in revealing the intricate properties of FeSC.
Naturally EXAFS studies using single crystals are also going to be
important, in addition to the angular dependent local information
provided by the polarization dependent EXAFS measurements
\cite{loc_str_G4}.

\begin{figure}
\includegraphics[width=10.0 cm]{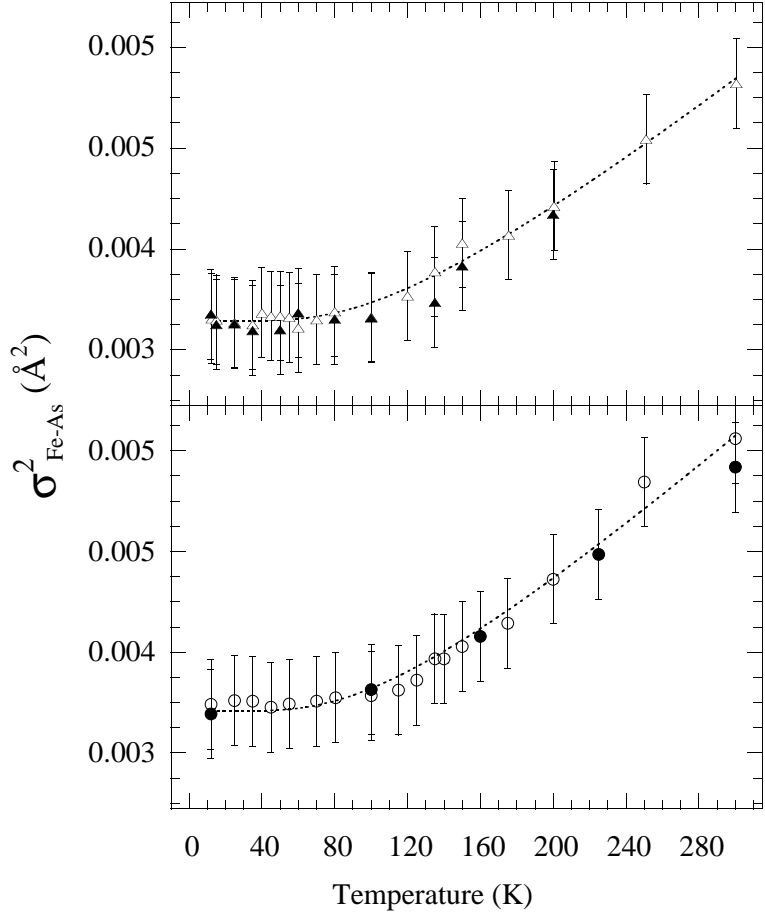}
\caption{\label{fig:epsart} 
Temperature dependence of the Fe-As MSRD (symbols) for the
NdFeAsO$_{1-x}$F$_x$ ($x$=0.0, 0.05, 0.15 and 0.18) system.  Upper
panel shows the data on the superconducting samples (open and filled
symbols corresponds to $x$=0.18 and 0.15 samples respectively), while
the lower panel shows the data on the non-superconducting samples
(open and filled symbols corresponds to x=0.0 and 0.05 samples
respectively).  The dotted lines represent the correlated Einstein
model fit to the data.  The Einstein-temperatures, $\Theta_{E}$, are
348$\pm$12 K and 326$\pm$12 K respectively for the superconducting
($x$=0.18) and non-superconducting ($x$=0.0) samples.  The temperature
independent $\sigma_{0}^{2}$ is found to have similar value
($\sim$0.00073) in both the cases.}
\end{figure}

Figure 4 shows mean square relative displacements ($\sigma^{2}$) of
the Fe-As pair, describing distance-distance correlation function
(correlated Debye Waller factors).  The MSRD is a sum of temperature
independent ($\sigma_{0}^{2}$) and temperature dependent terms \cite{Konings}, i.e.,
\begin{equation}
\sigma_{Fe-As}^{2}=\sigma_{0}^{2}+\sigma_{Fe-As}^{2}(T) \nonumber
\end{equation}
In several cases, the temperature dependent term can be described by 
the correlated Einstein-model \cite{Konings,book2},
\begin{equation}   
\sigma_{Fe-As}^{2}(T) = \frac{\hbar}{2 \mu_{Fe-As}\omega_E} coth(\frac{\hbar\omega_E}{2k_BT}),\nonumber
\end{equation}
where $\mu_{Fe-As}$ is the reduced mass of the Fe-As bond and $\omega_E$ is the 
Einstein-frequency. The related Einstein-temperature can be obtained
from the expression $\Theta_E=\hbar\omega_E/k_B$.
In the present case, temperature dependence of $\sigma_{Fe-As}^{2}$ is found to 
follow the correlated-Einstein model. Description of $\sigma_{Fe-As}^{2}$ using 
this model yield $\Theta_{E}$ values to be 348$\pm$12 K and 326$\pm$12 K respectively for
the $x$=0.18 and the $x$=0.0 samples. Notice that the uncertainity in $\Theta_E$ given 
above are obtained from the least-square fit considering the error bars on the data poitns.    
The obtained Einstein-frequencies are in the range of phonon-frequencies observed
in Raman studies for the modes involving arsenic and iron atoms
\cite{Raman_Nd111,Raman_Nd111-no-strong-e-p}.  In addition, different
Einstein frequencies ($\omega_{E}$) indicate different local force
constants (k=$\mu_{Fe-As}\omega_{E}^2$, where k is the effective force
constant) for the Fe-As
bonds in the superconducting and non-superconducting samples.  The
calculated local force constants for the Fe-As bonds are $\sim$6.65
eV/\AA$^2$ and $\sim$5.85 eV/\AA$^2$ respectively for the
superconducting and non superconducting samples.  Thus the Fe-As
bondlength seems to get harder in the superconducting regime.  This
observation is an indication of non-negligible role of the lattice
modes in the superconductivity of these materials.  

 Comparing the already available local structural data on different
 oxypnictides
 \cite{La1111-EXAFS,Sm1111-EXAFS,La1111-EXAFS-2,Ba122-EXAFS,iadecola_epl,Ba122-PDF}
 one can see that the Fe-As bond in these systems shows only little
 changes with doping and temperature.  An EXAFS study on a series of
 oxypnictides \cite{iadecola_epl} have shown that the Fe-As bondlength
 and the related MSRDs hardly show any change with the varying rare
 earth size, consistent with the strong covalent nature of the Fe-As
 bonds.  On the other hand, earlier studies on F doped La-1111
 \cite{La1111-EXAFS} and Sm-1111 \cite{Sm1111-EXAFS} compounds have
 shown presence of an anomaly in the temperature dependence of the
 Fe-As MSRDs.  But these anomalies were seen to be very weak comparing
 to that observed for the cuprates \cite{loc_str_G4}.  Indeed another
 study on superconducting La-1111 system \cite{La1111-EXAFS-2} did not
 indicate any such anomaly.  Such anomalies were also not seen in the
 K doped Ba-122 system \cite{Ba122-EXAFS} and F doped Ce-1111 system
 \cite{Ce1111-EXAFS}.  From the results presented in Figure 4, for the
 F-doped Nd-1111 system, there is no evidence for a significant
 anomaly in the MSRD of Fe-As bonds associated with the
 superconducting transition.  Here, we make an explicit comparison of
 the Fe-As MSRDs of the superconducting Sm-1111 and Nd-1111 samples to
 underline that the Fe-As MSRDs tend to change around the
 superconducting transition temperature, albeit the anomalies are
 indeed weak in compare to what has been seen in the copper oxide
 superconductors \cite{loc_str_G4}.

\begin{figure}
\includegraphics[width=10.0 cm]{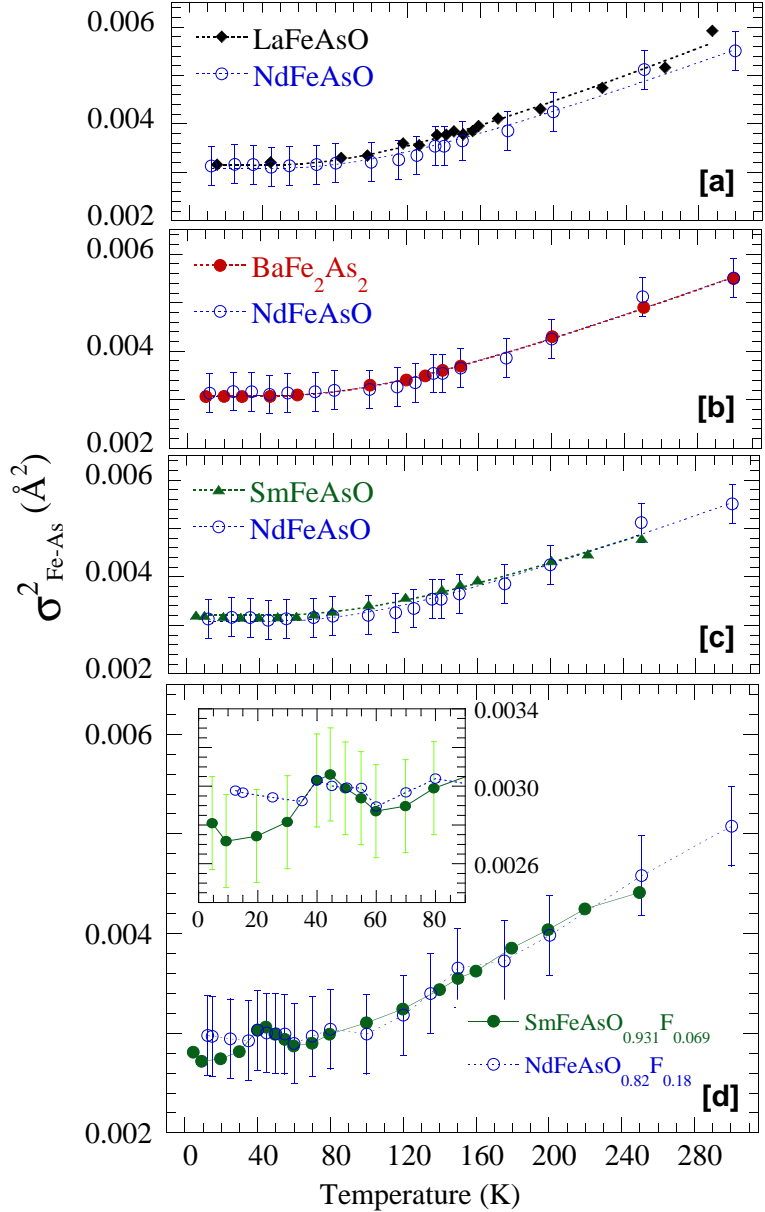}
\caption{\label{fig:epsart} 
Temperature dependence of the Fe-As MSRDs for the NdFeAsO system (open
circles in all panels) is compared with that of the LaFeAsO \cite{La1111-EXAFS-2} 
(filled diamonds (a)), SmFeAsO \cite{Sm1111-EXAFS} (filled triangles, (b)) 
and the BaFe$_2$As$_2$ \cite{Ba122-EXAFS} (filled circles, (c)).  
For facilitating a better comparison, the LaFeAsO \cite{La1111-EXAFS-2}, 
BaFe$_2$As$_2$ \cite{Ba122-EXAFS} and SmFeAsO \cite{Sm1111-EXAFS} data 
are scaled respectively by 0.0004, 0.0008 and 0.0017.  
Panel (d) presents the temperature dependence of the Fe-As MSRD for the 
F doped superconducting Nd-1111 and Sm-1111 \cite{Sm1111-EXAFS} (with the
latter scaled by 0.0012).  Overall temperature dependence of the two
superconducting systems are identical.  Inset in panel (d) shows a
zoom around the superconducting transition temperature.  The
error-bars in the inset are identical as that given in Ref.
\cite{Sm1111-EXAFS}.}
\end{figure}

Figure 5 shows temperature dependence of the MSRDs of the Fe-As bond
for the NdFeAsO, compared with that of the LaFeAsO
\cite{La1111-EXAFS-2}, SmFeAsO \cite{Sm1111-EXAFS} and BaFe$_2$As$_2$
\cite{Ba122-EXAFS}.  In all the cases, the MSRD data are extracted
from the arsenic $K$-edge EXAFS. Correlated-Einstein model fit to the
data is also indicated in the figure (as dotted lines).  From Fig.  5,
it can be seen that the Fe-As MSRDs show almost identical temperature
dependence for these different oxypnictide systems.  Although the
temperature dependence looks very similar, there seems to be a
systematic change in the Einstein-temperature.  The $\Theta_E$ values are
respectively, 316$\pm$5 K, 318$\pm$10 K, 326$\pm$ 12 K, and 328$\pm$
12 K for the LaFeAsO, BaFe$_2$As$_2$, NdFeAsO and SmFeAsO.  
There is a clear increase in the force-constant of the
Fe-As bonds, with increasing RE size in the ``1111'' series.  Optical
studies using single crystals of the ``1111'' series reveal that the
Fe-As stretching mode in these oxypnictides shows substantial
hardening with the rare-earth size, being harder for the NdFeAsO and
SmFeAsO showing higher T$_{c}$ in compare to the LaFeAsO
\cite{Raman-RE}.  The E$_u$ mode frequency for the BaFe$_2$As$_2$ \cite{IR-Ba122}  
is also found to be similar to that of the ``1111'' series
\cite{Raman-RE}.  In the ``1111'' family, the changing rare-earth
ionic size leads to a change in the pnictogen height above the
Fe-plane \cite{iadecola_epl} and thus a different interlayer atomic
correlations \cite{XANES_REOFeAs,XANES_As_REOFeAs}.  Local structural
studies using atomic pair distribution function analysis on
Ba$_{1-x}$K$_x$Fe$_2$As$_2$ show that, although the changes in the
Fe-As bonds are minimum, the FeAs$_4$ tetrahedra show a systematic
change with K doping \cite{Ba122-PDF}.  Indeed a small change
in the pnictogen height is known to have significant effect on the
electronic properties of the system through changing degeneracy
between different Fe 3d bands (in particular between the
$d_{x^{2}-y^{2}}$ and $d_{xz}/d_{yz}$), with a direct implication on
the magnetic structure and superconductivity
\cite{Mazin_Johannes,KurokiPRL}.  Very recent photo-emission studies
have showed this pnictogen height dependent changes in the electronic
structure \cite{ARPES-pnictogen-height} confirming the importance of
even very small local structural changes in determining the properties
of the FeSCs.

To further enlighten possible anomaly around the superconducting
transition temperature, we have made an explicit comparison between
the temperature dependent Fe-As MSRDs for the F doped highest T$_c$
Nd-1111 and Sm-1111 \cite{Sm1111-EXAFS} superconductors (see, e.g.,
Fig.  5[d]).  The inset in figure  5[d] is a zoomed view around T$_c$.
The error-bars given here are identical as given in Ref.  \cite{Sm1111-EXAFS}.  
For the NdFeAsO$_{0.82}$F$_{0.18}$ sample, there is a weak anomaly in the
temperature dependence of the Fe-As MSRDs similar to that observed in
the superconducting SmFeAsO$_{0.931}$F$_{0.069}$ \cite{Sm1111-EXAFS}.
However, the anomaly is well within the experimental uncertainties.
This situation is different from that of cuprates \cite{loc_str_G4}
where several measurements showed clear lattice anomalies associated
with the superconducting transition.  Although more experiments are
needed, the weakness of the anomalies does not imply that lattice
fluctuations have minor role in the Fe-based superconductors.  Indeed,
the an empirical relationship between the superconducting transition
temperature and the FeAs$_4$ tetrahedra in FeSCs
\cite{CHLee,anion_ht_takano,egami} clearly establishes the role of
local lattice in the superconducting properties.  Local structural
studies also underline the same point \cite{Ba122-PDF,iadecola_epl}.
The increased force constant of the Fe-As bonds in the superconducting
sample compared to the parent compound in the Nd-1111 system indicate
the importance of lattice dynamics in determining the properties.
Systematic changes in the Fe-As force constant within the different
iron-pnictide compounds, as revealed by the present study, further
support this.

\section{CONCLUSIONS} 

In conclusion, we have carried out systematic temperature-dependent
local structural studies of the NdFeAsO$_{1-x}$F$_{x}$ ($x$=0.0, 0.05,
0.15 and 0.18) high temperature pnictide superconductor 
using arsenic $K$-edge EXAFS. In all the samples, the
temperature dependence of the mean-square relative-displacements
(MSRDs) of the Fe-As bonds are found to follow the correlated-Einstein
model.  However, the Einstein-frequency for the superconducting sample is 
higher than that of the parent compound, indicating a hardening of the Fe-As 
bond in the former compared to the latter.  
The over-all temperature dependence of the MSRDs of the Fe-As bond seems 
to be similar in the NdFeAsO, SmFeAsO, LaFeAsO and BaFe$_2$As$_2$, 
but with a systematic variation of the corresponding force constants.  
The changes occurring in the FeAs$_4$ tetraheadra, together with the 
coupling between the active layer and the spacer layer, may account 
for the changes in properties of different iron oxypnictide superconductors.

\begin{acknowledgments}
The authors thank Drs.  Sergey Nikitenko and Miguel Silveria of BM26A,
ESRF, Grenoble for their cooperation in the EXAFS measurements.  Research at 
Pavia university  is supported by the CARIPLO foundation (Project No. 2009- 2540 
‘Chemical control and doping effects in pnictide high temperature superconductors’).
\end{acknowledgments}

\end{document}